\documentclass[12pt]{article}
\usepackage{setspace}
\usepackage{amsmath,amssymb,mathrsfs}
\usepackage{color}
\usepackage{hyperref}
\begin{document}
\setlength{\topmargin}{-1cm} 
\setlength{\oddsidemargin}{-0.25cm}
\setlength{\evensidemargin}{0cm}
\newcommand{\e}{\epsilon}
\newcommand{\beq}{\begin{equation}}
\newcommand{\eeq}[1]{\label{#1}\end{equation}}
\newcommand{\bea}{\begin{eqnarray}}
\newcommand{\eea}[1]{\label{#1}\end{eqnarray}}
\renewcommand{\Im}{{\rm Im}\,}
\renewcommand{\Re}{{\rm Re}\,}
\newcommand{\diag}{{\rm diag} \, }
\newcommand{\Tr}{{\rm Tr}\,}
\def\draftnote#1{{\color{red} #1}}
\def\bldraft#1{{\color{blue} #1}}
\def\n{n \cdot v}
\def\ni{n\cdot v_I}
\setcounter{tocdepth}{2}\begin{titlepage}
\begin{center}

\vskip 4 cm
{\Large \bf {THE CHIRAL LIMIT OF FULLY NONLINEAR MINIMAL MASSIVE GRAVITY}}

\vskip 1 cm

{Massimo Porrati \footnote{E-mail: \href{mailto:mp9@nyu.edu}{mp9@nyu.edu}} 
and Xilin Sheng \footnote{E-mail: \href{mailto:xs2597@nyu.edu}{xs2597@nyu.edu}}}
\vskip .75 cm
{\em Center for Cosmology and Particle Physics, \\ Department of Physics, New York University, \\ 726 Broadway, New York, NY 10003, USA}
\end{center}
\vskip 1.25 cm
\begin{abstract}
\noindent  
We study 3D Anti de Sitter Minimal Massive Gravity in two regimes: a) at the chiral limit where one of the boundary Brown-Henneaux central charges vanishes and two modes become null and b) in the regime that one of the two charges is much larger than the other. At the chiral point, we  go beyond the known free-theory analysis to prove that these modes
  decouple completely also in the full MMG Lagrangian and field equations. We also use the full action to ascertain if the interacting theory becomes infinitely strongly coupled in the neighborhood of the chiral limit, where the theory is anyway not unitary. We show that this is not the case at tree level but that a strong coupling pathology appears in loops, starting at a loop level determined by the number of external legs in the bulk Feynman diagrams. 
 Finally, we show that there is no strong coupling problem in the second regime where, instead, the two boundary fields and the
 field propagating in the bulk decouple from one another.
\end{abstract}
\end{titlepage}
\newpage
\section{Introduction}\label{s1}
Gravity in three spacetime dimension is interesting both as a simplified model of gravity possessing some of the most interesting features of the more complex four dimensional theory --e.g. black holes for negative cosmological constant 
$\Lambda=-1/l^2$-- and, also for $\Lambda<0$, as a holographic dual of two dimensional conformal
field theories (CFTs). 3D pure gravity with negative $\Lambda$ 
does not propagate any local degree of freedom in the bulk but it possesses 
boundary excitations that give rise to a boundary Virasoro$\times$Virasoro algebra with a classical 
nonzero central charge $c=3l/2G_3$ ($G_3$=Newton's constant)~\cite{BH}. The simplest formulation of this theory is in terms of 
an $SL(2,\mathbb{R})\times SL(2,\mathbb{R})$ Chern-Simons theory~\cite{CS}. The Chern-Simons theory depends on the two
couplings  of the two gauge factors, but the equations of motion for any nonzero coupling are always equivalent to
Einstein's equation for pure gravity.

The $SL(2,\mathbb{R})\times SL(2,\mathbb{R})$  theory does change if the zero-torsion constraint is added to the Chern-Simons 
theory. The resulting theory is Topologically Massive Gravity (TMG)~\cite{TMG}, which possesses two independent 
central charges, $c_+,c_-$. Generically, TMG contains ghosts whenever 
the Newton constant's sign is chosen to give positive boundary Brown-Henneaux~\cite{BH} central charges, but ghosts
appear to be absent at the chiral point, where either $c_+$ or $c_-$ vanish, provided that the theory is restricted to the
singlet subsector of the Virasoro algebra with vanishing central charge~\cite{TMG1,TMG2}.

The holographic duality~\cite{m,gkp,w1,agmoo} between gravity in $AdS_3$ and 2D conformal field theories (CFTs) can be
checked in great details thanks to the rich symmetry structure of 2D CFTs. One interesting question one can ask is: what is the
bulk dual of a renormalization group (RG) flow from some high energy (UV) CFT with central charges $(c_+\gg 1, c_-\gg 1)$ to
a chiral CFT with either $c_+=0$ or $c_-=0$? 
A candidate for such bulk dual is TMG, which in fact has been studied as a holographic 
model of a chiral ($c_--c_+ \neq 0$) RG flow in~\cite{bch}. The problem with such bulk dual is that TMG propagates ghosts away
from the strict chiral limit. A proper bulk dual must be unitary even away from the chiral point. The simplest such model is 
minimal massive gravity~\cite{MMG} (MMG), which can be ghost and tachyon free for generic positive central charges.

For generic values of $c_+>0,c_->0$, 
MMG propagates two boundary and one (chiral) bulk degrees of freedom. At the chiral point where either $c_-=0$ or $c_+=0$,  the bulk degree of freedom decouples together with one of the two boundary ones. At the level of the free, quadratic action, such
decoupling was verified in~\cite{Tek}. That paper worked in the metric formalism and showed that the Ostrogradsky energy of the
unwanted modes vanishes at the chiral points. 
This is of course a harbinger of potential problems away from the
strict chiral limit, because a theory with vanishing energy (or, equivalently, vanishing kinetic term) can become infinitely  coupled near the limit. If the kinetic term of a field is infinitesimal, e.g. $O(\eta)$ for $\eta\rightarrow 0$, interaction terms can become $O(\eta^{-1})$. This problem is distinct from another known problem of MMG, namely, that it not unitary near the chiral point~\cite{redux}.

To understand if MMG possesses a strong coupling pathology we must go beyond the free, quadratic action and study the fully
nonlinear MMG. This is done in this letter, which begins in section~\ref{s2} with a review of the MMG action written in 
first-order formalism, its equations of motion, and its linearization around an $AdS_3$ background.
Section~\ref{s3} describes the chiral limit of the quadratic part of the MMG action, where one of the two central charges, which we choose to be $c_-$ for definiteness, vanishes. Section~\ref{s4} is the first truly new part of this work. It contains a detailed study of the full MMG action near the
chiral limit. An appropriate rescaling of the fluctuations around the classical $AdS_3$ background allows to recover in a very simple
fashion the results of ~\cite{Tek} and show how interactions behave as a function of a constant, $\eta$, which parametrizes 
 the distance from the chiral limit. The results of section~\ref{s4} are used in section~\ref{s5} to show that the boundary mode 
 that survives in the chiral limit, called $f^a_-$, is not strongly coupled at bulk tree level. This is sufficient to establish the consistency of MMG as
 a holographic (non-unitary) 
 dual of a large-central-charge CFT, because in the large central charge regime (a.k.a. ``large-N'') the holographic
 dual is a classical gravity bulk action. Section~\ref{s5} also shows that certain loop diagrams containing only $f^a_-$ external legs 
 {\em can} 
 be strongly interacting at loop level. The minimum number of loops for this pathology to appear is related to the number of 
 external $f^a_-$ legs by an equation given in the section. Section~\ref{s5} concludes with a brief discussion of the 
 interpretation of, and possible solutions to, the strong coupling pathology.
 Section~\ref{s6} studies another region of MMG, in which the two boundary central charges $c_+$ and $c_- <c_+$ 
 are very different, 
 but still in the region allowed by bulk unitarity. Precisely, we study the region where $(c_+ - c_-/ c_+ + c_-)=1-\epsilon$,
 with $0<\epsilon\ll 1$. We find that no strong coupling pathology exists there. The two boundary degrees of freedom and the bulk
  degree of freedom are healthy and propagate in the
 limit $(c_+ - c_-/ c_+ + c_-)\rightarrow 1$, but decouple from one another, and one of them becomes free.

 \section{The MMG action in first order formalism}\label{s2}
The Lagrangian density of MMG in first order formalism is the three form
\beq
    \begin{split}
       16\pi G_3 \mathcal{L}_{MMG} = -\sigma e^a \wedge R(\omega)_a + \frac{\Lambda_0}{6} \epsilon_{abc} e^a \wedge e^b \wedge e^c +  &h^a \wedge T(\omega)_a + \frac{1}{2 \mu} (\omega^a \wedge dw_a + \frac{1}{3} \epsilon_{abc} \omega^a \wedge \omega^b \wedge \omega^c) \\ + \frac{\alpha}{2} \epsilon_{abc} & e^a \wedge h^b \wedge h^c .
    \end{split}
\eeq{eq:MMG Lagrangian (before linearization)}
Here $e^a$ is the dreibein, $\omega^a$ is the spin-connection and $h^a$ is a non-propagating degree of freedom~\cite{MMG}. 
$\Lambda_0$ is a parameter. Furthermore, $\mu$ is a mass parameter while $\sigma=\pm 1$. The
sign can be chosen freely without loss of generality and we set $\sigma = +1$ in this letter. 
The torsion tensor $T(\omega)_a$ and Riemann tensor $R(\omega)_a$ are~\cite{MMG}
\beq
    T(\omega)_a = de_a + \epsilon_{abc} \omega^a \wedge e^b , \quad
    R(\omega)_a = d\omega_a + \frac{1}{2} \epsilon_{abc} \omega^a \wedge \omega^b .
\eeq{R and T tensor def}
The MMG Lagrangian density differs from the TMG one because of a single term, namely $16\pi G_3\mathcal{L}_{MMG} = 16\pi G_3\mathcal{L}_{TMG} + \frac{\alpha}{2} \epsilon_{abc} e^a \wedge h^b \wedge h^c$~\cite{MMG}. The extra term transforms the
Lagrange multiplier $h^a$ into a non-propagating degree of freedom appearing quadratically in the action. 
The Lagrange multiplier appear linearly in the TMG action, 
so its equations of motion enforce the zero torsion constraint $T(\omega)_a=0$. The MMG action is instead quadratic in $h^a$, so
the equations of motion of this non-propagating degree of freedom remove it from the action without imposing any constraint. 
Two interesting limit of MMG are $\alpha=0$ and $\alpha=\infty$. In the first limit MMG reduces to TMG; in the second it reduces 
to the $SL(2,\mathbb{R})\times SL(2,\mathbb{R})$ Chern-Simons theory formulation of pure gravity~\cite{CS}. The first
property follows from the definition of MMG while the second is seen by solving the $h^a$ equations of motion, which give a
term proportional to $ \alpha^{-1} T^2$, which vanishes in the limit $\alpha\rightarrow \infty$.

The classical field equations for MMG are~\cite{MMG}:
\bea
   0 &=& T(\omega)_a + \alpha \epsilon_{abc} e^b \wedge h^c , \\ \notag
   0 &=& R(\omega)_a + \mu(1 + \alpha \sigma) \epsilon_{abc} e^b \wedge h^c  , \\ \notag
    0 &=& Dh_a + \frac{\alpha}{2} \epsilon_{abc} h^b \wedge h^c + \frac{\Lambda_0}{2} \epsilon_{abc}e^b \wedge e^c + \mu \sigma(1 + \alpha\sigma) \epsilon_{abc} e^b \wedge h^c  ,
\eea{eq:MMG field eqs (before linearization)}
 where $D = d + \epsilon_{abc} \omega^b$ is the covariant derivative with respect to the spin-connection field $\omega^a$.

The MMG action can be linearized around an $AdS_3$ background that solves the classical equations of motion by expanding
the one form fields $e^a$, $\omega^a$ and $h^a$ into a background, $\bar{e}^a, \, \bar{\omega}^a,\, \bar{h}^a\propto \bar{e}^a$, and three propagating modes, $f_+^a$, $f_-^a$ and $p^a$~\cite{MMG}, according to
\bea
    e^a &=& \bar{e}^a + l (f_+^a + f_-^a) + \frac{1}{\mu (1 - 2C)} p^a  , \\ \notag
    \omega^a &=& \bar{\omega}^a -\alpha C \mu \bar{e}^a + (1 - \alpha C \mu l)f_+^a - (1 + \alpha C \mu l)f_-^a + \frac{M + \alpha \mu (C - 1)}{\mu (1 - 2C)} p^a , \\ \notag
    h^a &=& C \mu \bar{e}^a + C \mu l f_+^a + C\mu l f_-^a + \frac{\mu (1 - C)}{\mu (1 - 2C)} p^a ,
\eea{eq:MMG linearization methods}
where
\begin{center}
    \begin{align} \label{eq:MMG coefficient variables}
    \sigma \frac{\Lambda_0}{\mu^2} &\equiv  \frac{\Lambda}{\mu^2} -\alpha(1 + \sigma\alpha)C^2, &  
    C &\equiv -\frac{\Lambda + \alpha\Lambda_0}{2 \mu^2 (1 + \sigma\alpha)^2}, \\ \notag
    M &\equiv \mu[\sigma(1 + \sigma\alpha) - \alpha C] , &
    1 -2C &\equiv \frac{(Ml)^2 - 1}{(1 + \sigma\alpha)^2(l \mu)^2}  .
    \end{align}
\end{center}
The new parameter $\Lambda$ is the cosmological constant of the $AdS_3$ background, so $\Lambda = -\frac{1}{l^2}$ and 
$l$ is the $AdS_3$ radius~\cite{MMG}.
 The identity relating $C$ to $M$ in eq.~\eqref{eq:MMG coefficient variables} immediately implies (when $\sigma = +1$)
\beq
    [C^2 \alpha^2 + 2\alpha C + 2C]l^2 \mu^2 - 1 = 0 ,
\eeq{eq: Alternative form of the identity between C and M}
which is crucially important when deriving the linearized MMG Lagrangian
and analyzing the region near the chiral limit.

The boundary central charge for MMG is \cite{MMG}
\beq
    c_{\pm} \propto \Sigma_{\pm} \equiv \sigma \pm \frac{1}{\mu l} + \alpha C .
\eeq{eq:MMG central charge}
When $\sigma=+1$ eqs.~\eqref{eq: Alternative form of the identity between C and M} and \eqref{eq:MMG central charge} further yield
\beq
    \Sigma_{-}\Sigma_{+} = 1 - 2C .
\eeq{eq: 1 -2C parametrization}

The MMG Lagrangian expanded to quadratic order around the $AdS_3$ background (with $\sigma = +1$) is~\cite{MMG}: 
\beq
    \begin{split}
  16\pi G_3      \mathcal{L}^{(2)}_{MMG} = -l \Sigma_{-} [f_+^a \wedge \bar{D}{f_+}_a &+ l^{-1} \epsilon_{abc} \bar{e}^a \wedge f_+^b \wedge f_+^c] + l \Sigma_{+} [f_-^a \wedge \bar{D}{f_-}_a - l^{-1} \epsilon_{abc} \bar{e}^a \wedge f_-^b \wedge f_-^c]\\
        &+ \frac{1}{2 \mu \Sigma_{-}\Sigma_{+}} [p^a \wedge \bar{D}{p}_a + M \epsilon_{abc} \bar{e}^a \wedge p^b \wedge p^c] ,
    \end{split}
\eeq{eq:quadratic L_MMG chiral sigma = 1}
where $\bar{D} = d + \epsilon_{abc} \bar{\omega}^a$.

The equations of motion obeyed by $\mathcal{L}^{(2)}_{MMG}$ are~\cite{MMG}:
\beq
    \bar{D}{f_+}_a + l^{-1}\epsilon_{abc} \bar{e}^b \wedge f_+^c = 0 , \quad
    \bar{D}{f_-}_a - l^{-1}\epsilon_{abc} \bar{e}^b \wedge f_-^c = 0  , \quad
    \bar{D}{p}_a + M\epsilon_{abc} \bar{e}^b \wedge p^c = 0 .
\eeq{eq:chiral qudratic field eqs sigma = 1}

\subsection{Alternative formulation} \label{s2.1}
The MMG action can be written in a second-order metric formulation.  Expanding around the $AdS_3$ background with metric
\beq
    ds^2 \equiv \bar{g}_{\mu\nu} dx^\mu dx^\nu = -\frac{1}{\Lambda}(-\cosh^2 \rho dt^2 + d\rho^2 + \sinh^2 \rho d\phi^2),
\eeq{eq:AdS_3 metric}
one obtains an action, whose quadratic form was given in~\cite{Tek} in the gauge $\bar{\nabla}_{\mu}h^{\mu \nu} = 0$.
Writing $h = \bar{g}^{\mu \nu}h_{\mu \nu}$ the result is
\beq
    S = -\frac{1}{4 \pi G_3} \int d^3 x \sqrt{-\bar{g}}[(\sigma - \frac{\gamma\Lambda}{2 \mu^2})[\bar{\nabla}_{\alpha}h_{\mu \nu} \bar{\nabla}^{\alpha}h^{\mu \nu} + 2\Lambda h_{\mu \nu}h^{\mu \nu}] + \frac{1}{\mu}\epsilon^{\alpha \beta}_{\mu}\bar{\nabla}_{\alpha}h_{\mu \nu}(\bar{\Box} - 2\Lambda)h_{\beta \nu}] , 
\eeq{eq:Alternative MMG action}
The definition of $\Lambda$ and $\sigma$ is as before, while  $\gamma$ is a dimensionless parameter. 
The boundary central charges are~\cite{Tek}:
\beq
    c_L = \frac{3l}{2 G_3}(\sigma + \frac{\gamma}{2 l^2 \mu^2} - \frac{1}{l \mu}), \quad   c_R = \frac{3l}{2 G_3}(\sigma + \frac{\gamma}{2 l^2 \mu^2} + \frac{1}{l \mu}).
\eeq{eq: alternative central charge}
Furthermore, $h_{\mu \nu}$ is expanded into three propagating modes $h_{\mu \nu}$ = $h^M_{\mu \nu}$ + $h^R_{\mu \nu}$ + $h^L_{\mu \nu}$ and their field equations are ~\cite{Tek}:
\beq
    \epsilon^{\alpha \beta}_{\mu} \bar{\nabla}_{\alpha} h^M_{\beta \nu} + \mu(\sigma - \frac{\gamma\Lambda}{2 \mu^2})h^M_{\beta \nu} = 0 , \quad
    \epsilon^{\alpha \beta}_{\mu} \bar{\nabla}_{\alpha} h^L_{\beta \nu} + l^{-1}h^L_{\beta \nu} = 0 ,  \quad
    \epsilon^{\alpha \beta}_{\mu} \bar{\nabla}_{\alpha} h^R_{\beta \nu} -l^{-1}h^R_{\beta \nu} = 0 .
\eeq{eq: alternative field eqs}

 \section{The chiral limit}\label{s3}
We are only concerned here with the case where $\sigma = +1$ and we choose $c_- = 0$ for the chiral limit.  Analogous 
statements can be made for the other limit, $c_+=0$ by replacing $l \rightarrow -l$ and $f_+^a \leftrightarrow -f_-^a$ etc. in
 the formulas presented in this paper. After substituting $c_- = 0$ into eq.~\eqref{eq:MMG coefficient variables}, we find the
  values for the parameters of the theory at the chiral limit:
\beq
    C = \frac{1}{2} , \quad
    M = l^{-1 } , \quad
    \alpha = \frac{2}{l \mu} - 2, \quad
    \Sigma_-  = 0 , \quad
    \Sigma_+ = 2 .
\eeq{eq:MMG chiral limit coe}
It is clear that at the chiral limit any term involving the $p$ mode will have infinite coupling strength. This strong coupling can be real or just an effect of a bad normalization. To see which case it is, we rescale the mode $p$ by defining $p^a = (1 - 2C)q^a = \Sigma_+\Sigma_- q^a$. Then, eq.~\eqref{eq:quadratic L_MMG chiral sigma = 1} becomes
\beq
  16\pi G_3  \mathcal{L}^{(2)}_{MMG} = \frac{2}{\mu} [f_-^a \wedge \bar{D}{f_-}_a - l^{-1} \epsilon_{abc} \bar{e}^a \wedge f_-^b \wedge f_-^c]
    .
\eeq{eq:chiral quadratic L_MMG sigma = 1}
Notice that the two modes $f_+^a$ and $q^a$ have disappeared completely from the action, so the only propagating degree of freedom is the chiral boundary graviton $f_-^a$ and the equations of motion for $f_+^a$ and $q^a$ are not determined by the 
action. If we derive the equations of motion before taking the chiral limit, we do get that three field equations in \eqref{eq:chiral qudratic field eqs sigma = 1} become:
\beq
    \bar{D}{f_+}_a + l^{-1}\epsilon_{abc} \bar{e}^b \wedge f_+^c = 0 , \quad
    \bar{D}{f_-}_a - l^{-1}\epsilon_{abc} \bar{e}^b \wedge f_-^c = 0  , \quad
    \bar{D}{q}_a + l^{-1} \epsilon_{abc} \bar{e}^b \wedge q^c = 0 .
\eeq{eq:chiral qudratic field eqs sigma = 1 at chiral limit}
The important point here is that the equations of motion for $f_+^a$ and $q^a$ decouple from $f_-^a$, which is necessary for the
consistency of the chiral limit.

Our results are in agreement with the analysis of the chiral limit of MMG done in~\cite{Tek} which we breifly summarize 
in~\ref{s3.1}. The first-order formalism used in this letter makes it easy to extends the analysis, done in~\cite{Tek}
 only at quadratic order and in a Hamiltonian formalism, to the full interacting action. 
 The linearization plus rescaling of the action done here is, in our opinion, the simplest way to check that the unwanted excitations $f_+^a,q^a$ do in fact decouple in the full theory.

\subsection{Alternative formulation at the chiral limit} \label{s3.1}
We choose $c_L = 0$ for the alternative formulation at the chiral limit. Then, the MMG action becomes:
\beq
    S = -\frac{1}{2 \pi l \mu G_3} \int d^3 x \epsilon^{\alpha \beta}_{\mu} \bar{\nabla}_{\alpha}{h^R}^{\mu \nu} \bar{\Box} h^R_{\beta \nu} + \frac{2}{l^2} \epsilon^{\alpha \beta}_{\mu} (\bar{\nabla}_{\alpha} {h^R}^{\mu \nu}) h^R_{\beta \nu}.
\eeq{eq: alternative MMG action at chiral limit}
The field equations obeyed the three propagating modes at the chiral limit are:
\beq
    \epsilon^{\alpha \beta}_{\mu} \bar{\nabla}_{\alpha} h^M_{\beta \nu} + l^{-1}h^M_{\beta \nu} = 0 , \quad
    \epsilon^{\alpha \beta}_{\mu} \bar{\nabla}_{\alpha} h^L_{\beta \nu} + l^{-1}h^L_{\beta \nu} = 0 ,  \quad
    \epsilon^{\alpha \beta}_{\mu} \bar{\nabla}_{\alpha} h^R_{\beta \nu} -l^{-1}h^R_{\beta \nu} = 0 .
\eeq{eq: alternative field eqs at chiral limit}
It is clear that $h^M_{\mu \nu}$, $h^L_{\mu \nu}$ and $h^R_{\mu \nu}$ correspond to $q^a$, $f_+$ and $f_-$ respectively.

\section{The near-chiral limit}\label{s4}
Even chiral TMG~\cite{TMG1,TMG2} appears to be consistent and ghost free in the singlet sector of the ``$-$'' boundary Virasoro algebra --which has $c_-=0$ so it does admit a singlet unitary representation. On the other hand, TMG does contain ghosts for any $c_->0$, so it is not 
consistent near the chiral point $c_-=0$. MMG, on the other hand, makes sense also for $c_->0$, at least at the level of the 
noninteracting quadratic action. To investigate whether MMG is consistent and computable near the chiral limit is the aim of this section and it is the main new result of this letter.

We begin by writing in full the MMG action expanded around the $AdS_3$ background. In first order formalism, it contains cubic terms besides the quadratic ones.
The cubic-order MMG Lagrangian for $\sigma = +1$ is 
\beq
    \begin{split}
    16\pi G_3    \mathcal{L}^{(3)}_{MMG} = 
        -\frac{2l}{3}\Sigma_{-} & \epsilon_{abc} f_+^a \wedge f_+^b \wedge f_+^c
        + \frac{lM + 1}{2 \mu \Sigma_{-}\Sigma_{+}} \epsilon_{abc} f_+^a \wedge p^b \wedge p^c 
        - \frac{2l}{3}\Sigma_{+} \epsilon_{abc} f_-^a \wedge f_-^b \wedge f_-^c\\
        + \frac{lM - 1}{2 \mu \Sigma_{-}\Sigma_{+}} & \epsilon_{abc}f_-^a \wedge p^b \wedge p^c
        + \frac{(2+ \alpha)\mu + 2M}{6 \mu^2 (\Sigma_{-}\Sigma_{+})^2} \epsilon_{abc} p^a \wedge p^b \wedge p^c .
    \end{split}
\eeq{eq:cubic order MMG when signma = 1}

To analyze the region near the chiral limit we take the central charge $c_-=\eta \ll 1$ and expand in powers of $\eta$:
\beq
    c_- \propto \Sigma_{-} = 1 - \frac{1}{\mu l} + \alpha C = \eta .
\eeq{eq:method for near chiral}
Next we use eqs.~\eqref{eq: Alternative form of the identity between C and M} and~\eqref{eq:method for near chiral}, to find 
that, near the chiral limit, $C$, $\alpha$ and $M$ have the following expansion in powers of $\eta$
\beq
    C = \frac{1}{2} - \frac{\eta}{l \mu}, \quad  
    \alpha = \frac{2l\mu \eta + 2 - 2l\mu}{l\mu - 2\eta}, \quad
    M = \frac{\mu[(-l\mu + 4 - \frac{2}{l \mu})\eta - 1]}{2 \eta - l\mu} .
\eeq{eq:coefficient variables when near the chiral limit}
We see that after rescaling the $p^a$ mode as $p^a = (1 - 2C)q^a = \Sigma_+\Sigma_- q^a$ and keeping only the lowest 
order of $\eta$ in the numerator and the zeroth order in the denominator, the cubic-order MMG Lagrangian near the chiral limit is 
\beq
    \begin{split} 
   16\pi G_3     \mathcal{L}^{(3)}_{MMG} = 
        -\frac{2l \eta}{3} &\epsilon_{abc}  f_+^a \wedge f_+^b \wedge f_+^c
        + \frac{2\eta}{l\mu^2} \epsilon_{abc} f_+^a \wedge q^b \wedge q^c \\
        - (\frac{4}{3 \mu} + \frac{2 l \eta}{3}) &\epsilon_{abc} f_-^a \wedge f_-^b \wedge f_-^c
        + (\frac{1}{\mu} - \frac{4}{l\mu^2} + \frac{4}{l^2 \mu^3})\eta^2 \epsilon_{abc}f_-^a \wedge q^b \wedge q^c \\
        +& \frac{4\eta}{3l^2\mu^3} \epsilon_{abc} q^a \wedge q^b \wedge q^c .
    \end{split}
\eeq{eq:cubic order MMG near the chiral}
Two  points to keep in mind here are: a) all cubic terms except the third one, which contains only $f_-^a$, vanish at $\eta=0$;
b) There exist no terms in~\eqref{eq:cubic order MMG near the chiral} which contain two $f_-^a$'s and either one $q^a$ or
one $f_+^a$. These two points are crucial to show the consistency of the tree level expansion of MMG near the chiral limit, which we shall prove in the next section.

At the chiral point $\eta=0$ the full Lagrangian is 
\beq
16\pi G_3 \mathcal{L}_{MMG}=\frac{2}{\mu} [f_-^a \wedge \bar{D}{f_-}_a - l^{-1} \epsilon_{abc} \bar{e}^a \wedge f_-^b \wedge f_-^c]
- \frac{4}{3 \mu} \epsilon_{abc} f_-^a \wedge f_-^b \wedge f_-^c .
\eeq{full chiral action}
As for the quadratic action~\ref{eq:chiral quadratic L_MMG sigma = 1}, the full action also depends only on $f_-^a$. The 
additional modes $f_+^a$, $q^a$ have completely decoupled.

The chiral-point limit of the three field equations obeyed by the full Lagrangian is
\bea
    0 &=& \bar{D}{f_+}_a + l^{-1}\epsilon_{abc} \bar{e}^b \wedge f_+^c + \epsilon_{abc}f_+^b \wedge f_+^c - \frac{1}{l \mu} \epsilon_{abc}q^b \wedge q^c , \\ \notag
    0 &=& \bar{D}{f_-}_a - l^{-1}\epsilon_{abc} \bar{e}^b \wedge f_-^c - \epsilon_{abc} f_-^b \wedge f_-^c , \\ \notag
    0 &=& \bar{D}{q}_a + l^{-1}\epsilon_{abc} \bar{e}^b \wedge q^c + \frac{2}{l\mu} \epsilon_{abc} q^b \wedge q^c + 2\epsilon_{abc} f_+^b \wedge q^c .
\eea{eq: field eqs for 2nd plus 3rd MMG action after rescaling}
As in~\eqref{eq:chiral qudratic field eqs sigma = 1 at chiral limit}, the modes $f_+^a, q^a$ decouple from $f_-^a$ 
not only in the action  but also in the equations of motion.

 \section{Unitarity and strong coupling}\label{s5}
 
 MMG is not unitary near the chiral limit $c_-=0$ for $\eta \geq 0$~\cite{redux}. It is still interesting to study this region in view of
 possible applications where the holographic dual is non unitary, as in statistical mechanics models and in view of other 
 considerations explained at the end of this section. Besides 
 that, it is instructive to check the region since $c_-=0$ (equivalently, $c_+=0$) is
 a potentially dangerous point, because it is there that the kinetic term of the two modes $f_+$ and $p^a$ vanishes. We saw that
 the rescaling of $p^a=\Sigma_+ \Sigma_-q^q$ makes it manifest that at $\eta=0$ the whole action for $f_+$ and $q^a$ vanish. So these modes disappear from the theory without producing infinite strong coupling pathologies. 
 The behavior of the theory near $\eta=0$ is different. There, all modes propagate and the presence of one power of $\eta$ in the
 kinetic term of $q^a$ and $f_+^a$ makes it possible that strong coupling pathologies are present. 
 In fact a simple power counting argument shows that they do exist, but they are not present at tree level.
 
 The first feature that we need to prove that MMG is (classically) not strongly coupled is that the only field that can appear as an external state
 in a Feynman diagram expansion of the action is $f_-^a$. This is the field that remains physical at the chiral point. After all, 
 even in TMG at the chiral point, there exist non-unitary modes that are nonsinglets of the chiral left Virasoro algebra~\cite{GKP,TMG2}. Certainly, there could  be a strongly coupled sector in the near-chiral MMG, but it would be
  harmless if it was sequestered from the physical sector, whose propagating field is $f_-^a$ only.
  The previous remark shows that at tree level MMG is safe, because a tree-level diagram in which some of the dangerous modes $q^a, f_+^a$ propagate and with only $f_-^a$s as external legs
  must contain at least two vertices with  two $f_-^a$s and either one $q^a$ or one $f_+^a$. But 
  eq.~\eqref{eq:cubic order MMG near the chiral} shows that such vertices do not exist. 
  
  Beyond tree level, the theory still makes sense up to a loop order that depends on the number of external $f_-^a$ legs.
  A standard Feynman diagram counting shows that the number of vertices, $V$, external $f_-^a$ legs, $E$, 
  propagators $P$, loops $L$, and leading powers of $\eta$, $H$,  are related by
  \beq
  2P+E=3V, \quad L=1+P-V , \quad H= V+E-P .
  \eeq{mass1}
 These equations give $H=1+E-L$, so a diagram with $E$ external $f_-^a$ legs exhibits a string coupling pathology --i.e. negative
 powers of $\eta$-- only at $L> 1+E$ loops.
 
 It is our opinion that neither the loop divergences of MMG, nor indeed its lack of unitarity near the chiral limit  necessarily signal a 
 lethal pathology of the theory, but only of its classical 
 action. First of all, if we use MMG as a bulk dual of a large-$N$ boundary 2D CFT, it is the classical limit that matter in the 
 description of the large-$N$ limit. Finite $N$ corrections involve genuine bulk quantum gravity effects. So the chiral point, where 
 one of the central charges is very small, is not under perturbative control. On the same vein, at finite
 $N$ (e.g. finite Newton constant) the action is not power-counting 
 renormalizable so it requires counterterms not present in the classical
 action. Those counterterms could cancel the negative-$H$ divergences order-by-order in the loop expansion. 
 Whether this is indeed possible is an interesting question, but quite out of reach with perturbative bulk gravity techniques. 
 So, we move instead to another region, where both $c_+$ and $c_-$ are positive {\em and large} but where the ratio
 \beq
 {c_+ -c_- \over c_+ + c_-}\lesssim 1.
 \eeq{ineq}
 This is the limit where the ratio approaches 1 from below.

 \section{Left-right asymmetric regime} \label{s6}
 This is the regime where the two boundary central charges are ``as chiral as possible,'' that is when they differ by a large amount
 while lying in the unitary region of MMG parameters. In this regime it is convenient to use the parametrization of~\cite{redux}, which defines 
\beq
    x = 2C , \quad y = \frac{2}{l\mu} .
\eeq{new parametrization}
The central charges become ~\cite{redux}
\beq
    c_\pm = \frac{3l}{2 G_3} \Sigma_\pm = \frac{3l}{2 G_3}(1 + \frac{\alpha x}{2} \pm \frac{y}{2}) ,
\eeq{new parametrization: central charges and the identity}

and eq.~\eqref{eq: 1 -2C parametrization} is transformed to 
\beq
    \Sigma_+\Sigma_- = 1 - x .
\eeq{new parametrization: 1 - 2C}
Combining eqs.~(\ref{new parametrization: central charges and the identity},\ref{eq: 1 -2C parametrization}) we get the equation
of a hyperbola in the $x,y$ plane~\cite{redux} 
\beq
y^2 =\alpha^2 x^2 + 4(1+\alpha) x .
\eeq{hyp}
While minimal massive gravity is not unitary near a chiral point, it is unitary in a region where the two central charges take
very different values, namely where
\beq
\frac{c_+ - c_-}{c_+ + c_-} =  {y\over 2+ \alpha x} = 1- \epsilon, \qquad 0< \epsilon \ll 1.
\eeq{chir-reg}

Since $\alpha<0$, $y>0$~\cite{redux}, this limit is reached for negative values of $x$ and $\alpha x\rightarrow \infty$, where 
\beq
y= \alpha x  + {2(1+\alpha) \over \alpha } + O(1/x) .
\eeq{lim-y}
Here we are interested in studying minimal massive gravity, in which $\alpha$ is finite and nonzero. So, we study the limit
$x\rightarrow -\infty$, $\alpha=$constant. 
Substituting eq.~\eqref{lim-y} into $c_\pm$ we find
\beq
c_+=  \frac{3l}{2 G_3}\left(2 + \alpha x+ \frac{1}{\alpha}\right), \quad
c_-=  \frac{3l}{2 G_3}\left(-\frac{1}{\alpha}\right) .
\eeq{l-r-charges}
Both charges are positive and $c_+$ diverges in the limit.

The quadratic Lagrangian written in the new parametrization in terms of $\alpha,x,y$ is
\beq
    \begin{split}
   16\pi G_3     \mathcal{L}^{(2)}_{MMG} = -l (1 + \frac{\alpha x}{2} - \frac{y}{2}) [&f_+^a \wedge \bar{D}{f_+}_a + l^{-1} \epsilon_{abc} \bar{e}^a \wedge f_+^b \wedge f_+^c] \\
        + l (1 + \frac{\alpha x}{2} + \frac{y}{2}) [&f_-^a \wedge \bar{D}{f_-}_a - l^{-1} \epsilon_{abc} \bar{e}^a \wedge f_-^b \wedge f_-^c]\\
        + \frac{ly}{4 (1 - x)} &p^a \wedge \bar{D}{p}_a + \frac{(2 + 2\alpha - \alpha x)}{4(1 - x)} \epsilon_{abc} \bar{e}^a \wedge p^b \wedge p^c ,
    \end{split}
\eeq{new parametrization: quadratic L_MMG}
We see that in the limit $x\rightarrow -\infty$, $\alpha=\mbox{constant}<0$ the kinetic terms of $f^a_+$ and $p^a$ remain constant
while the kinetic term of $f_-^a$ diverges proportionally to $\alpha x $. The mass term of $p^a$ approaches a finite value. 
The cubic part of the Lagrangian is
\beq
    \begin{split}
    16\pi G_3    \mathcal{L}^{(3)}_{MMG} = 
        l\Big[-\frac{2 + \alpha x - y}{3} & \epsilon_{abc} f_+^a \wedge f_+^b \wedge f_+^c
        + \frac{2 + 2\alpha - \alpha x + y}{4(1 - x)} \epsilon_{abc} f_+^a \wedge p^b \wedge p^c \\
        - \frac{2 + \alpha x + y}{3} & \epsilon_{abc} f_-^a \wedge f_-^b \wedge f_-^c
        + \frac{2 + 2\alpha - \alpha x - y}{4 (1- x)}  \epsilon_{abc}f_-^a \wedge p^b \wedge p^c \\
        + \frac{(4 + 3\alpha - \alpha x)y}{12 (1 - x)^2} & \epsilon_{abc} p^a \wedge p^b \wedge p^c\Big] .
    \end{split}
\eeq{new parametrization: cubic L_MMG}
All interactions are finite in the limit, except the $(f_-^a)^3$ term, which diverges proportionally to $\alpha x$. The interaction term
$f_+^a (p^a)^2$ vanishes. By redefining $f_-^a= (\alpha x)^{-1/2} \phi^a_-$ and taking the limit $x\rightarrow -\infty$ we finally 
arrive at the following simple result:
\begin{enumerate}
\item All interaction terms with at least one $\phi_-^a$ vanish, so $\phi_-^a$ becomes a decoupled free field.
\item $f_+^a$ and $p^a$ they become two decoupled self-interacting fields, without any interaction with
each other of $\phi_-^a$. 
\end{enumerate}

We conclude that in extreme asymmetric limit, which is well inside the unitary region of parameters for MMG~\cite{redux}, the
theory exhibits no pathologies and if fact becomes quite simple. 
The limit can be taken while keeping both $c_+$ and $c_-$ large, so it is well within the validity regime of a semiclassical 
holographic interpretation. It is certainly interesting to see if the simplification in the 3D bulk theory corresponds to a simplification 
of the dual boundary conformal field theory, but we leave this question for future study.

 \subsection*{Acknowledgements} 
M.P. would like to thank the TH Department at CERN for its kind hospitality during completion of this work. 
M.P. is supported in part by NSF grant PHY-2210349. 
 

\end{document}